\documentstyle[12pt]{article}

\newcommand{\ab}{\bar{a}}
\newcommand{\cb}{\bar{c}}
\newcommand{\db}{\bar{d}}
\newcommand{\lb}{\bar{l}}
\newcommand{\mb}{\bar{m}}
\newcommand{\pb}{\bar{p}}
\newcommand{\xb}{\bar{x}}
\newcommand{\lambdab}{\bar{\lambda}}
\newcommand{\yyb}{\bar{Y}}
\newcommand{\hhb}{\bar{H}}
\newcommand{\llb}{\bar{L}}
\def\case#1#2{{\textstyle{#1\over #2}}}

\title{
Connection between type A and E factorizations and construction of satellite
algebras}
\author{A. Del Sol Mesa $^{a,}$\thanks{E-mail address:
delsol@gredos.cnb.uam.es}\ ,
C. Quesne $^{b,}$\thanks{Directeur de recherches FNRS; E-mail address:
cquesne@ulb.ac.be}\\
$^a$ {\small Protein Design Group, CNB-CSIC. Campus Universidad Aut\'onoma,}\\
{\small Cantoblanco, Madrid M-28049, Spain}\\
$^b$ {\small Physique Nucl\'eaire Th\'eorique et Physique
Math\'ematique,  Universit\'e Libre de Bruxelles,} \\
{\small Campus de la Plaine CP229, Boulevard~du Triomphe, B-1050 Brussels,
Belgium}}
\date{ }
\begin{document}
\baselineskip=20pt plus 1pt minus 1pt
%%%%%%%%%%%%%%%%%%%%%%%%%%%%%%%%%%%%%%%%%%%%%%%%%%%%%%%%%%
\maketitle
%
%======================================================================
%
\newpage
\noindent
{\bf Abstract.} Recently, we introduced a new class of symmetry algebras, called
satellite algebras, which connect with one another wavefunctions belonging to
different potentials of a given family, and corresponding to different energy
eigenvalues. Here the role of the factorization method in the construction
of such
algebras is investigated. A general procedure for determining an so(2,2) or
so(2,1)
satellite algebra for all the Hamiltonians that admit a type E factorization is
proposed. Such a procedure is based on the known relationship between type
A and E
factorizations, combined with an algebraization similar to that used in the
construction of potential algebras. It is illustrated with the examples of the
generalized Morse potential, the Rosen-Morse potential, the Kepler problem in a
space of constant negative curvature, and, in each case, the conserved
quantity is
identified. It should be stressed that the method proposed is fairly
general since
the other factorization types may be considered as limiting cases of type A or E
factorizations.\par
%
%=========================================================================
%
\newpage
\section{Introduction}

Lie algebraic techniques have proved very useful in explaining the exact
solvability
of quantum mechanical problems~\cite{miller}. Such techniques arose from
the factorization method, introduced by Schr\"odinger~\cite{schrodinger} and later
developed by Infeld and Hull~\cite{infeld51}.\par
%
%------------------------------------------------------------------------
%
More recent developments in algebraic methods, such as the introduction of the
potential algebra concept~\cite{alhassid,wu,frank,barut87} and the SUSYQM
superalgebra scheme~\cite{witten} for shape-invariant
potentials~\cite{gendenshtein}, also heavily rely on the factorization
method (see
e.g.~\cite{barut87,montemayor}). All these approaches allow one to connect with
one another wavefunctions $\psi^{(m)}(x)$ corresponding to the same energy
eigenvalue, but to different potentials $V^{(m)}(x)$, $m = 0$, 1, 2,
\ldots,~$n-1$, of
a given family, which may be called {\em satellite potentials}. In the
factorization
method, the ladder operators connecting $\psi^{(m)}(x)$ to $\psi^{(m+1)}(x)$ or
$\psi^{(m-1)}(x)$ are $m$-dependent. In the potential algebra approach, this
$m$-dependence is eliminated by introducing some auxiliary variables, so
that the
resulting operators become the generators of some Lie algebra.  The latter is
compact or noncompact according to whether $n$ is finite or infinite. In the
SUSYQM approach, in contrast, the same elimination is performed by transforming
the ladder operators into supercharge ones and by introducing a supersymmetric
Hamiltonian, thereby giving rise to an su(1/1) superalgebra.\par
%
%-----------------------------------------------------------------------
%
In a recent work~\cite{delsol}, we introduced a new class of symmetry algebras,
which may be called {\em satellite algebras}. They are similar to the potential
algebras in the sense that they also depend upon some auxiliary variables and
connect among themselves wavefunctions belonging to different satellite
potentials. However, they are more general than the potential algebras,
because the
related wavefunctions correspond to different energy eigenvalues. There actually
exists a conserved quantity, different from the energy, which is the
eigenvalue of
the algebra Casimir operator.\par
%
%---------------------------------------------------------------------
%
In the case studied in ref.~\cite{delsol}, which is that of the generalized
Morse
potential (GMP)~\cite{deng} (related to the Manning-Rosen~\cite{manning} or
Eckart~\cite{eckart} potential), the conserved quantity is some combination
of the
potential parameters. This is an interesting property of the GMP satellite
algebra,
which may find applications in molecular physics. It is indeed well
known~\cite{nicholls} that when analysing electromagnetic transitions
between rovibrational bands in diatomic molecules, the initial and final
electronic states are in general different and therefore give rise to different
vibrational potentials, which should be taken into account in the calculation of
Frank-Condon factors. It was suggested by Ley-Koo~\cite{leykoo} that finding
an algebra that both changes the potential and the vibrational state could be
useful in this context. If we identify the initial and final potentials with GMP
satellite ones and the initial and final vibrational states with some
eigenstates
 of the latter, the GMP satellite algebra turns out to be a good candidate
for such an
algebra.\par
%
%-----------------------------------------------------------------------
%
Since this shows that the new class of satellite algebras may be physically
relevant, it is worth exploring it in more detail. In ref.~\cite{delsol},
the GMP
satellite algebra so(2,2) was constructed in an indirect way by connecting the
corresponding Schr\"odinger equation with either the Laplace equation on the
hyperboloid or the Schr\"odinger equation for the P\"oschl-Teller
potential, then
transferring the known so(2,2) symmetry algebra of the latter to the former. The
relation between this procedure and the factorization method, although implicit,
was left untouched.\par
%
%----------------------------------------------------------------------
%
The purpose of the present paper is to investigate the role of the factorization
method in the construction of satellite algebras. We shall devise a general
procedure for determining an so(2,2) or so(2,1) satellite algebra for all the
Hamiltonians that admit a type E factorization. Such a procedure is based
upon the
known relationship between type A and E factorizations~\cite{infeld51}, combined
with an algebraization similar to that used in the construction of potential
algebras~\cite{alhassid,wu,frank,barut87}. It should be noted that our
procedure is
fairly general since the other factorization types (B, C, D), and F may be
considered
as limiting forms of A and E, respectively.\par
%
%------------------------------------------------------------------------
%
The general method proposed here will allow us to recover and generalize the
results previously obtained for the GMP~\cite{delsol}. Various examples will be
presented for illustrative purposes, but it is obvious that detailed numerical
applications of each of them do not come within the scope of the present
paper and
are left for future work.\par
%
%-----------------------------------------------------------------------
%
This paper is organized as follows. The factorization method is briefly
reviewed in
section~2 and used in section~3 to provide a general construction method of
satellite algebras. In sections~4, 5, and 6, the latter is illustrated by
considering
the cases of the GMP, the Rosen-Morse potential, and the Kepler problem in
a space
of constant negative curvature, respectively. Finally, section~7 contains the
conclusion.\par
%
%=======================================================================
%
\section{The factorization method}
{}Following Infeld and Hull~\cite{infeld51}, the linear second-order
differential
equation
\begin{equation}
  \frac{d^2 y}{dx^2} + r(x,m) y + \lambda y = 0  \label{eq:IH-eq}
\end{equation}
where $m$ is a nonnegative integer and $\lambda$ the eigenvalue to be
determined,
can be factorized if it can be replaced by each of the following two equations:
\begin{eqnarray}
  H^+ (m+1)  H^-(m+1) y(\lambda,m) & = & [\lambda - L(m+1)] y(\lambda,m)
         \label{eq:IH-fact1} \\
  H^- (m)  H^+(m) y(\lambda,m) & = & [\lambda - L(m)] y(\lambda,m)
         \label{eq:IH-fact2}
\end{eqnarray}
where
\begin{equation}
  H^{\pm}(m) = \pm \frac{d}{dx} + k(x,m).  \label{eq:ladder}
\end{equation}
Here $k(x,m)$ is some function of $x$ and of the parameter $m$, and $L(m)$ is an
$m$-dependent real number.\par
%
%-----------------------------------------------------------------------
%
If equation~(\ref{eq:IH-eq}) can be factorized and $y(\lambda,m)$ is one of its
solutions, then $H^-(m+1)$ and $H^+(m)$ act as ladder operators, i.e., they
give rise
to other solutions
\begin{equation}
  y(\lambda,m+1) = H^-(m+1) y(\lambda,m) \qquad y(\lambda,m-1) = H^+(m)
  y(\lambda,m)
\end{equation}
corresponding to the same $\lambda$, but different $m$ values. Moreover, the
operators $H^-(m)$ and $H^+(m)$ are formally mutually adjoint.\par
%
%----------------------------------------------------------------------
%
In practical cases, it can be checked that if $y(\lambda,m)$ is a
square-integrable
solution of Eqs.~(\ref{eq:IH-fact1}) and~(\ref{eq:IH-fact2}), and $L(m)$ is an
increasing (resp.\ decreasing) function of $m$, then the operator
$H^-(m+1)$ (resp.\
$H^+(m)$) yields a function $y(\lambda,m+1)$ (resp.\ $y(\lambda,m-1)$) that is
also square integrable.\par
%
%-----------------------------------------------------------------------
%
When $L(m)$ is an increasing (resp.\ decreasing) function of $m$, the problem is
said to be of class~I (resp.~II). A necessary condition for square-integrable
solutions is then that $\lambda = \lambda_l = L(l+1)$ (resp.\ $\lambda =
\lambda_l
= L(l)$), where $l$ is an integer and $m=0$, 1, \ldots,~$l$ (resp.\ $m=l$,
$l+1$,~\ldots).\par
%
%-----------------------------------------------------------------------
%
Square-integrable solutions that are also normalized are denoted by $Y_l^m$. In
addition to equations (\ref{eq:IH-fact1}) and (\ref{eq:IH-fact2}), they
satisfy the
relations
\begin{equation}
  H^-(m+1) Y_l^m = [\lambda - L(m+1)]^{1/2} Y_l^{m+1} \qquad
  H^+(m) Y_l^m = [\lambda - L(m)]^{1/2} Y_l^{m-1}.  \label{eq:ladder-action-E}
\end{equation}
\par
%
%---------------------------------------------------------------------
%
The possible factorizations types can be found  by inserting
equation~(\ref{eq:ladder}) into equations (\ref{eq:IH-fact1})
and~(\ref{eq:IH-fact2}), comparing the results with
equation~(\ref{eq:IH-eq}), and
eliminating the function $r(x,m)$. This leads to a differential-difference
equation
for $k(x,m)$, which was shown by Infeld and Hull~\cite{infeld51} to have six
different nontrivial types of solutions, denoted by the letters A, B, C, D,
E, F. From
$k(x,m)$, it is then possible to find $r(x,m)$ and $L(m)$.\par
%
%----------------------------------------------------------------------
%
{}For type A and E factorizations, to be considered in the remainder of
this paper,
$r(x,m)$, $k(x,m)$, and $L(m)$ are given in terms of some constants $a$,
$c$, $d$,
$p$, $q$ by
\begin{eqnarray}
  r(x,m) & = & \frac{a^2 (m+c) (m+c+1) + d^2 + 2ad \left(m+c+\frac{1}{2}\right)
          \cos[a(x+p)]}{\sin^2[a(x+p)]}  \label{eq:r-A} \\
  k(x,m) & = & (m+c) a \cot[a(x+p)] + \frac{d}{\sin[a(x+p)]}  \label{eq:k-A} \\
  L(m) & = & a^2 (m+c)^2  \label{eq:L-A}
\end{eqnarray}
and
\begin{eqnarray}
  r(x,m) & = & - \frac{m (m+1) a^2}{\sin^2[a(x+p)]} - 2aq \cot[a(x+p)]
\label{eq:r-E}
          \\
  k(x,m) & = & ma \cot[a(x+p)] + \frac{q}{m}  \label{eq:k-E} \\
  L(m) & = & a^2 m^2 - \frac{q^2}{m^2}  \label{eq:L-E}
\end{eqnarray}
respectively.
%
%========================================================================
%
\section{General construction method of satellite algebras}
\setcounter{equation}{0}
Let us consider the most general second-order differential equation admitting a
type E factorization. From equations (\ref{eq:IH-eq}) and~(\ref{eq:r-E}),
it is given
by
\begin{equation}
  \frac{d^2\psi}{dx^2} - \left\{\frac{m (m+1) a^2}{\sin^2[a(x+p)]} + 2aq
\cot[a(x+p)]
  \right\} \psi + \lambda \psi = 0  \label{eq:eq-E}
\end{equation}
where the normalized eigenfunctions $Y_l^m$ corresponding to the discrete
eigenvalues $\lambda = \lambda_l$ are denoted by $\psi$, $a$, $p$, $q$ are some
constants, and $l$, $m$ run over some nonnegative integers.
Equation~(\ref{eq:eq-E}) can be factorized as shown in equations
(\ref{eq:IH-fact1}), (\ref{eq:IH-fact2}), with $y(\lambda,m)$ replaced by
$\psi = Y_l^m$, and $k(x,m)$, $L(m)$ given in equations (\ref{eq:k-E}),
(\ref{eq:L-E}),
respectively. For definiteness' sake, in the following we shall restrict
ourselves to
class~I problems, but the results obtained can easily be accommodated to
class~II
problems by replacing $l+1$ by $l$, and changing the $m$ range accordingly.\par
%
%----------------------------------------------------------------------
%
In the ladder operator definition given in equations (\ref{eq:ladder}) and
(\ref{eq:k-E}), $m$ occurs in the denominator, so that an algebraization
along the
lines of references~\cite{alhassid,wu,frank, barut87} is not possible. To
carry out
such an algebraization, it is necessary to first transform the type E
factorizable
equation~(\ref{eq:eq-E}) into a type A one, which according to equations
(\ref{eq:IH-eq}) and (\ref{eq:r-A}) is given by
\begin{equation}
  \frac{d^2\chi}{dy^2} - \frac{\ab^2 (\mb+\cb) (\mb+\cb+1) + \db^2 + 2\ab\db
  (\mb+\cb+\frac{1}{2}) \cos[\ab (y+\pb)]} {\sin^2[\ab (y+\pb)]} \chi +
\lambdab \chi
  = 0.  \label{eq:eq-A}
\end{equation}
Here the variable $x$ is changed into $y$, a bar is put on top of all the
constants to
distinguish them from those used for type E factorization, and the normalized
eigenfunctions $\yyb_{\lb}^{\mb}$, corresponding to the eigenvalues $\lambdab =
\lambdab_{\lb}$, are denoted by $\chi$. From equations (\ref{eq:ladder}),
(\ref{eq:k-A}), and (\ref{eq:L-A}), it follows that the associated ladder
operators
$\hhb^{\pm}(\mb)$, which depend linearly on $\mb$, and the real constant
$\llb(\mb)$ can be written as
\begin{equation}
  \hhb^{\pm}(\mb) = \pm \frac{d}{dy} + (\mb+\cb) \ab \cot[\ab (y+\pb)] +
  \frac{\db}{\sin[\ab (y+\pb)]} \label{eq:ladder-A}
\end{equation}
and
\begin{equation}
  \llb(\mb) = \ab^2 (\mb+\cb)^2
\end{equation}
respectively.
\par
%
%------------------------------------------------------------------
%
By performing two successive changes of variable and of function,
\begin{eqnarray}
  z & = & \ln[\tan(a\xb/2)] \qquad \xb \equiv x + p \nonumber \\
  \psi(x) & = & [\sin(a\xb)]^{1/2} \phi(z) = (\cosh z)^{-1/2} \phi(z)
\label{eq:change1}
\end{eqnarray}
and
\begin{equation}
  y = {\rm i} z + \frac{\pi}{2} \qquad \phi(z) = \chi(y) \label{eq:change2}
\end{equation}
equation~(\ref{eq:eq-E}) is transformed into an equation of
type~(\ref{eq:eq-A}),
\begin{equation}
  \frac{d^2\chi}{dy^2} - \frac{(\lambda/a^2) - (1/4) + 2 {\rm i} (q/a) \cos y}
  {\sin^2y} \chi + \left(m + \case{1}{2}\right)^2  \chi  = 0.
\label{eq:eq-Abis}
\end{equation}
\par
%
%----------------------------------------------------------------------
%
Comparison between equations~(\ref{eq:eq-A}) and~(\ref{eq:eq-Abis}) shows that
the type A factorization constants $\ab$, $\cb$, $\db$, $\pb$, and
parameter $\mb$
are connected with the constants $a$, $q$, and the eigenvalue $\lambda$ of
type E
factorization through the relations
\begin{eqnarray}
  && \ab = 1 \qquad \pb = 0 \label{eq:link1} \\
  && \db \left(\mb + \cb + \frac{1}{2}\right) = {\rm i} \frac{q}{a}
\label{eq:link2} \\
  && (\mb + \cb) (\mb + \cb + 1) + \db^2 = \frac{\lambda}{a^2} - \frac{1}{4}.
         \label{eq:link3}
\end{eqnarray}
From equation~(\ref{eq:link2}), we get
\begin{equation}
  \mb + \cb + \frac{1}{2} = \frac{{\rm i} q}{a\db} \label{eq:mbar}
\end{equation}
and by substituting this expression into equation~(\ref{eq:link3}), the latter
becomes
\begin{equation}
  - \frac{q^2}{\db^2} + a^2 \db^2 = \lambda. \label{eq:lambda-1}
\end{equation}
\par
%
%----------------------------------------------------------------------
%
We know however that for a type E factorizable problem of class I, the
eigenvalue
$\lambda$ is given by
\begin{equation}
  \lambda = L(l+1) = a^2 (l+1)^2 - \frac{q^2}{(l+1)^2} \label{eq:lambda-2}
\end{equation}
where equation~(\ref{eq:L-E}) has been used. By equating the two
expressions~(\ref{eq:lambda-1}) and~(\ref{eq:lambda-2}) for $\lambda$, we obtain
a quadratic equation for $\db^2$ with two real solutions, $\db^2 = (l+1)^2$ and
$\db^2 = - q^2/[a^2 (l+1)^2]$. By using equation~(\ref{eq:mbar}) again, we
therefore
get four possible choices for $\db$ and $\mb + \cb + (1/2)$,
\begin{equation}
  \db = \epsilon (l+1) \qquad \mb + \cb + \frac{1}{2} = \frac{{\rm i}
\epsilon q}
  {a (l+1)} \label{eq:sol1}
\end{equation}
and
\begin{equation}
  \db = \frac{{\rm i} \epsilon q}{a (l+1)} \qquad \mb + \cb + \frac{1}{2} =
  \epsilon (l+1) \label{eq:sol2}
\end{equation}
where $\epsilon = \pm 1$ is a so far undetermined sign.\par
%
%-------------------------------------------------------------------
%
After inverting the transformations~(\ref{eq:change1})
and~(\ref{eq:change2}), and
taking equation~(\ref{eq:link1}) into account, the type A ladder
operators~(\ref{eq:ladder-A}) lead to ladder operators for the original
eigenfunctions $\psi$,
\begin{eqnarray}
  \tilde{H}^{\pm}(\mb) & \equiv & [\sin(a\xb)]^{1/2} \hhb^{\pm}(\mb)
          [\sin(a\xb)]^{-1/2} \nonumber \\
  & = & \mp {\rm i} \frac{\sin(a\xb)}{a} \frac{d}{d\xb} + {\rm i} \left(\mb
+ \cb \pm
          \frac{1}{2}\right) \cos(a\xb) + \db \sin(a\xb)
\end{eqnarray}
where any of the two substitutions defined in equations~(\ref{eq:sol1})
and~(\ref{eq:sol2}) may in principle be performed. We shall denote the resulting
operators by $\tilde{H}^{\pm}_1(\mb)$ and $\tilde{H}^{\pm}_2(\mb)$,
respectively.\par
%
%------------------------------------------------------------------
%
Such ladder operators can now be transformed into Lie algebra generators by
introducing two auxiliary variables $\xi$, $\eta \in [0, 2\pi)$, and extended
eigenfunctions defined by
\begin{equation}
  \Psi_{s,t}(x, \xi, \eta) = (2\pi)^{-1} e^{{\rm i} s \xi} \psi(x) e^{{\rm i} t \eta}
  \label{eq:extended}
\end{equation}
where
\begin{equation}
  s \equiv \frac{{\rm i} \epsilon q}{a (l+1)} \qquad t \equiv \epsilon (l+1).
  \label{eq:st}
\end{equation}
Since
\begin{equation}
  S_0 = - {\rm i} \frac{\partial}{\partial\xi} \qquad T_0 = - {\rm i}
  \frac{\partial}{\partial\eta} \label{eq:S_0}
\end{equation}
are such that
\begin{equation}
  S_0 \Psi_{s,t} = s \Psi_{s,t} \qquad T_0 \Psi_{s,t} = t \Psi_{s,t}
  \label{eq:S_0-action}
\end{equation}
we may replace $s$ and $t$ by $- {\rm i} \partial/\partial\xi$ and $- {\rm i}
\partial/\partial\eta$ when such operators act on the extended eigenfunctions,
respectively. By combining the transformations
\begin{equation}
  (- {\rm i}) e^{\pm {\rm i} \xi} \tilde{H}^{\mp}_1\left(\mb + \case{1}{2} \pm
  \case{1}{2}\right) \to S_{\pm} \qquad
  (- {\rm i}) e^{\pm {\rm i} \eta} \tilde{H}^{\mp}_2\left(\mb + \case{1}{2} \pm
  \case{1}{2}\right) \to T_{\pm}
\end{equation}
with these substitutions, we obtain
\begin{eqnarray}
  S_{\pm} & = & e^{\pm {\rm i} \xi} \left[\pm \frac{\sin(a\xb)}{a}
\frac{\partial}
        {\partial\xb} - {\rm i} \cos(a\xb) \frac{\partial}{\partial\xi} -
\sin(a\xb)
        \frac{\partial}{\partial\eta}\right] \label{eq:S} \\
  T_{\pm} & = & e^{\pm {\rm i} \eta} \left[\pm \frac{\sin(a\xb)}{a}
\frac{\partial}
        {\partial\xb} - {\rm i} \cos(a\xb) \frac{\partial}{\partial\eta} -
\sin(a\xb)
        \frac{\partial}{\partial\xi}\right]. \label{eq:T}
\end{eqnarray}
We note that $S_0$, $S_{\pm}$ and $T_0$, $T_{\pm}$ only differ by the
substitutions $\xi \leftrightarrow \eta$, $\partial/\partial\xi \leftrightarrow
\partial/\partial\eta$.\par
%
%-------------------------------------------------------------------
%
It is now straightforward to check that each set of generators $S_0$,
$S_+$, $S_-$
and $T_0$, $T_+$, $T_-$ satisfies the defining relations of $\mbox{\rm su(1,1)}
\simeq \mbox{\rm so(2,1)}$, e.g.,
\begin{equation}
  [S_0 , S_{\pm}] = \pm S_{\pm} \qquad [S_+, S_-] = - 2 S_0
\end{equation}
and that any generator of the first set commutes with any generator of the
second
one. Hence, the six operators generate an $\mbox{\rm so(2,2)} \simeq \mbox{\rm
su(1,1)} \oplus \mbox{\rm su(1,1)}$ Lie algebra.\par
%
%--------------------------------------------------------------------
%
Both Casimir operators
\begin{equation}
  C_s \equiv - S_+ S_- + S_0 (S_0-1) \qquad C_t \equiv - T_+ T_- + T_0 (T_0-1)
\end{equation}
are equal and given by
\begin{equation}
  C_s = C_t = C = \sin^2(a\xb) \left[\frac{1}{a^2}
\frac{\partial^2}{\partial\xb^2}
  - \frac{\partial^2}{\partial\xi^2} - \frac{\partial^2}{\partial\eta^2} -
2 {\rm i}
  \cot(a\xb) \frac{\partial^2}{\partial\xi\partial\eta}\right].
\end{equation}
Since from equations~(\ref{eq:lambda-2}) and~(\ref{eq:st}),
\begin{equation}
  s^2 + t^2 = \frac{\lambda}{a^2} \qquad s t = \frac{{\rm i} q}{a}
\end{equation}
the action of $C$ on the extended eigenfunctions~(\ref{eq:extended}) is given by
\begin{eqnarray}
  C \Psi_{s,t}(x, \xi, \eta)  & = & (2 \pi)^{-1} e^{{\rm i} (s\xi + t\eta)}
         \frac{\sin^2(a\xb)}{a^2} \left[\frac{d^2}{d\xb^2} - 2 a q \cot(a\xb) +
         \lambda\right] \psi(x) \nonumber \\
  & = & m (m+1) \Psi_{s,t}(x, \xi, \eta) \label{eq:casimir}
\end{eqnarray}
where in the last step use has been made of equation~(\ref{eq:eq-E}).\par
%
%---------------------------------------------------------------------
%
All the arguments presented so far have been rather formal. For
completeness' sake,
we also have to discuss the eigenfunction normalizability conditions, which are
known to play an important role in applying the factorization method. This
will be
considered for some examples in the next sections. At this stage, however,
we may
already note three important properties at the Lie algebra representation
level.\par
%
%-------------------------------------------------------------------
%
{}Firstly, it is clear that such representations will be nonunitary. The lack of
unitarity actually comes from the normalization change implied by the
transformation from $\psi(x)$ to $\phi(z)$ in equation~(\ref{eq:change1}).\par
%
%-------------------------------------------------------------------
%
Secondly, in practice we shall have to distinguish between trigonometric and
hyperbolic potentials, for which $a$ in equation~(\ref{eq:eq-E}) is real or
imaginary, respectively. In the former case, $s$, defined in
equation~(\ref{eq:st}),
turns out to be imaginary. This is incompatible with the eigenvalues of
$S_0$ in an
su(1,1) irreducible representation, which should differ from one another by some
real integer~\cite{barut65}. Hence, we are only left with the su(1,1) algebra
generated by $T_0$, $T_+$, and $T_-$. In the latter case, on the contrary,
by setting
$a = {\rm i} \alpha$ ($\alpha$ real), we find that
\begin{equation}
  s = \frac{\epsilon q}{\alpha (l+1)} \label{eq:s}
\end{equation}
so that both su(1,1) algebras may be considered. We shall concentrate on
this case
in the remainder of the present paper, and therefore replace
equations~(\ref{eq:eq-E}), (\ref{eq:S}), and~(\ref{eq:T}) by
\begin{equation}
  \frac{d^2\psi}{d\xb^2} - \left[\frac{m (m+1) \alpha^2}{\sinh^2(\alpha\xb)} +
  2 \alpha q \coth(\alpha \xb)\right] \psi + \lambda \psi = 0 \qquad (\xb
\equiv x +
  p) \label{eq:eq-Ebis}
\end{equation}
\begin{eqnarray}
  S_{\pm} & = & e^{\pm {\rm i} \xi} \left[\pm \frac{\sinh(\alpha\xb)}{\alpha}
        \frac{\partial}{\partial\xb} - {\rm i} \cosh(\alpha\xb) \frac{\partial}
        {\partial\xi} - {\rm i} \sinh(\alpha\xb)
\frac{\partial}{\partial\eta}\right]
        \label{eq:Sbis} \\
  T_{\pm} & = & e^{\pm {\rm i} \eta} \left[\pm \frac{\sinh(\alpha\xb)}{\alpha}
        \frac{\partial}{\partial\xb} - {\rm i} \cosh(\alpha\xb) \frac{\partial}
        {\partial\eta} - {\rm i} \sinh(\alpha\xb)
\frac{\partial}{\partial\xi}\right]
        \label{eq:Tbis}
\end{eqnarray}
respectively.\par
%
%---------------------------------------------------------------------
%
Thirdly, from equation~(\ref{eq:casimir}), we note that the so(2,2) irreps
may be
characterized by $m$, so that their basis functions may be denoted by
$\Psi_{s,t}^{(m)}(x,\xi,\eta)$. When acting on such functions, the generators
$S_{\pm}$ of the first su(1,1) algebra change $s$ into $s \pm 1$, while
leaving $t$
unchanged. In other words, the energy eigenvalue label $l$ and the potential
parameter $m$ do not change, but the other potential parameter $q$ becomes
$q' = q
\pm \epsilon \alpha (l+1)$. When considering instead the generators $T_{\pm}$ of
the second su(1,1) algebra, $s$ is left unchanged, while $t$ changes into
$t \pm 1$.
In this case, the potential parameter $m$ is still unchanged, but both $l$
and $q$
are changed into $l' = l \pm \epsilon$ and $q' = q (l + 1 \pm\epsilon)/(l+1)$,
respectively. It is therefore clear that the so(2,2) generators connect among
themselves eigenfunctions belonging to different satellite potentials and
different
energy eigenvalues. We conclude that any family of type E factorizable
Hamiltonians corresponding to hyperbolic potentials has an so(2,2) satellite
algebra.\par
%
%--------------------------------------------------------------------
%
It should be noted that there remains an undetermined sign $\epsilon$ in the
definitions~(\ref{eq:st}) and~(\ref{eq:s}) of $s$ and $t$. In all the examples
considered in the next sections, we have checked that apart from some irrelevant
phase factors, the results are independent of the choice made for
$\epsilon$. Hence,
in the remainder of this paper, we shall use the convention
\begin{equation}
  \epsilon = \frac{q}{|q|} \label{eq:epsilon}
\end{equation}
which provides the simplest link with the GMP analysis in
reference~\cite{delsol}.\par
%
%====================================================================
%
\section{The generalized Morse potential}
\setcounter{equation}{0}
As a first example, let us consider the GMP studied in references~\cite{delsol,
deng}. The corresponding Schr\"odinger equation is
\begin{equation}
  - \frac{\hbar^2}{2\mu} \frac{d^2\psi}{dr^2} + D \left(1 - \frac{b}{e^{ar} - 1}
  \right)^2 \psi - E \psi = 0 \qquad b = e^{ar_e} - 1 \label{eq:H-GMP}
\end{equation}
where $0 \le r < \infty$, and $D$, $b$, $a$ are some parameters regulating the
depth, position of the minimum $r_e$, and radius of the potential.\par
%
%-------------------------------------------------------------------
%
In terms of the parameters
\begin{equation}
  \alpha_n = \sqrt{k - \epsilon_n} \qquad \beta_n = \sqrt{\alpha_n^2 + k b
(b+2)}
  \qquad  m = \case{1}{2} \left(- 1 + \sqrt{1 + 4 k b^2}\right)
  \label{eq:parameters-GMP1}
\end{equation}
where
\begin{equation}
  k = \frac{2 \mu D}{a^2 \hbar^2} \qquad \epsilon_n = \frac{2 \mu E_n}
  {a^2 \hbar^2} \label{eq:parameters-GMP2}
\end{equation}
the energy eigenvalues and corresponding eigenfunctions are given by
\begin{equation}
  E_n = D - \frac{a^2 \hbar^2}{8\mu} \left(n + m + 1 - \frac{kb(b+2)}{n+m+1}
  \right)^2 \label{eq:energy-GMP}
\end{equation}
and
\begin{equation}
  \psi_n(r) = N_n y^{\alpha_n} (1+y)^{-\beta_n} {}_2F_{1}(-n, -n-2m-1; 2\alpha_n
  + 1; -y) \qquad y \equiv \left(e^{ar} - 1\right)^{-1} \label{eq:psi-GMP}
\end{equation}
where $n=0$, 1, \ldots,~$n_{max}$, $n_{max}$ is the largest integer smaller than
$\sqrt{kb(b+2)} - m - 1$, and $N_n$ is some normalization coefficient.\par
%
%------------------------------------------------------------------
%
Equation~(\ref{eq:H-GMP}) can be rewritten as a type E factorizable Hamiltonian
corresponding to a hyperbolic potential. By performing the change of
variable $\xb =
ar/2$, it indeed reduces to equation~(\ref{eq:eq-Ebis}), where
\begin{equation}
  \alpha = 1 \qquad q = - kb(b+2) \qquad \lambda = 4 (\epsilon - k) - 2kb(b+2)
  \label{eq:GMP-E}
\end{equation}
and $m$ is given by equation~(\ref{eq:parameters-GMP1}). The corresponding
$L(m) =
- m^2 - (q^2/m^2)$ is an increasing function of $m$, so that the GMP
problem is of
class I. Comparing then $\lambda = \lambda_l = L(l+1) = - (l+1)^2 -
q^2/(l+1)^2$ with
the expression for $\lambda$ resulting from
equations~(\ref{eq:parameters-GMP2}), (\ref{eq:energy-GMP}),
and~(\ref{eq:GMP-E}), we obtain the relation
\begin{equation}
  l = n + m
\end{equation}
between the eigenvalue labels $n$ and $l$, coming from the resolution of the
Schr\"odinger equation and the factorization method, respectively.\par
%
%--------------------------------------------------------------------
%
{}From such a relation, we find that $s$ and $t$, defined in
equations~(\ref{eq:st}),
(\ref{eq:s}), and~(\ref{eq:epsilon}), become
\begin{equation}
  s = \frac{kb(b+2)}{n+m+1} = \alpha_n + \beta_n \qquad t = -n-m-1 = \alpha_n
  - \beta_n
\end{equation}
and therefore correspond to the quantum numbers $m$ and $g$ of
reference~\cite{delsol}, respectively.\footnote{Note that the symbols $l$
and $m$
of reference~\cite{delsol} correspond to $m+1$ and $s$ in the present paper,
respectively.} Moreover, when rewritten in terms of the variables $y$,
$\xi$, $\eta$
(see equation~(\ref{eq:psi-GMP})), the so(2,2) generators $S_{\pm}$,
$T_{\pm}$ of
equations~(\ref{eq:Sbis}) and~(\ref{eq:Tbis}) are transformed into
\begin{eqnarray}
  S_{\pm} & = & e^{\pm{\rm i}\xi} \left(\mp \sqrt{y(y+1)}
\frac{\partial}{\partial
          y} - {\rm i} \frac{2y+1}{2\sqrt{y(y+1)}} \frac{\partial}{\partial
\xi} - {\rm i}
          \frac{1}{2\sqrt{y(y+1)}} \frac{\partial}{\partial\eta}\right) \\
  T_{\pm} & = & e^{\pm{\rm i}\eta} \left(\mp \sqrt{y(y+1)}
\frac{\partial}{\partial
          y} - {\rm i} \frac{2y+1}{2\sqrt{y(y+1)}}
\frac{\partial}{\partial\eta} - {\rm i}
          \frac{1}{2\sqrt{y(y+1)}} \frac{\partial}{\partial\xi}\right)
\end{eqnarray}
and therefore coincide with the operators $M^{\pm}$ and $- G^{\pm}$ of
reference~\cite{delsol}. From equation~(\ref{eq:parameters-GMP1}), it
follows that
the conserved quantity, given by the eigenvalue $m$ of the Casimir operator
$C$, is
here a combination of the potential parameters $kb^2$, or $Db^2/a^2$. The
operators
$S_{\pm}$ change $b$ into $b' = 2 (kb^2) b/(2kb^2 \mp tb)$ while leaving $n$
unchanged, whereas the operators $T_{\pm}$ change both $b$ and $n$ into $b' =
2tb/(2t \pm b \pm 2)$ and $n\mp1$, respectively. For both types of
operators, $k$
becomes $k' = kb^2/b'^2$.\par
%
%--------------------------------------------------------------------
%
We conclude that the results obtained in reference~\cite{delsol}, using
some ad hoc
arguments, are but special cases of the general formalism developed in the
present
paper. In the next two sections, we shall prove that other examples can be
treated
in a similar way.\par
%
%=====================================================================
%
\section{The Rosen-Morse potential}
\setcounter{equation}{0}
The Schr\"odinger equation for the Rosen-Morse potential~\cite{rosen} is
\begin{equation}
  - \frac{\hbar^2}{2\mu} \frac{d^2\psi}{dx^2} + \left[B \tanh(\alpha x) - C
  \mbox{\rm sech}^2(\alpha x)\right] \psi - E \psi = 0 \label{eq:H-RM}
\end{equation}
where $- \infty < x < + \infty$, $\alpha$ determines the radius of the
potential,
while $B$, $C$ regulate the position of its minimum $x_0 = - \alpha^{-1}
\tanh^{-1}
(B/2C)$ and its depth at the minimum $V(x_0) = - C - B^2/(4C)$, and are
restricted
by the condition $|B| < 2C$.\par
%
%---------------------------------------------------------------------
%
In terms of the parameters
\begin{equation}
  a_n = - \frac{\beta}{2b_n} \qquad b_n = \sqrt{\gamma + \case{1}{4}} - n -
  \case{1}{2} \qquad m = \case{1}{2} \left(- 1 + \sqrt{1 + 4\gamma}\right)
  \label{eq:parameters-RM1}
\end{equation}
where
\begin{equation}
  \beta = \frac{2\mu B}{\hbar^2 \alpha^2} \qquad \gamma = \frac{2\mu
  C}{\hbar^2 \alpha^2}
\end{equation}
the energy eigenvalues and corresponding eigenfunctions are given
by~\cite{rosen}
\begin{equation}
  E_n = - \frac{\hbar^2 \alpha^2}{2\mu} \left(a_n^2 + b_n^2\right)
  \label{eq:energy-RM}
\end{equation}
and
\begin{equation}
  \psi_n(x) = N_n e^{a_n \alpha x} [\cosh(\alpha x)]^{-b_n} {}_2F_1(-n,
2m-n+1; a_n
  + b_n + 1; y) \qquad y \equiv \case{1}{2} [1 + \tanh(\alpha x)]
\label{eq:psi-RM}
\end{equation}
where $n=0$, 1, \ldots,~$n_{max}$, $n_{max}$ is the largest integer smaller than
$m - \sqrt{|\beta|/2}$, and $N_n$ is some normalization
coefficient~\cite{nieto}.\par
%
%--------------------------------------------------------------------
%
Equation~(\ref{eq:H-RM}) can be rewritten in the form~(\ref{eq:eq-Ebis}) by
setting
\begin{equation}
  p = {\rm i} \frac{\pi}{2\alpha} \qquad q = \frac{1}{2} \alpha \beta \qquad
  \lambda = \frac{2\mu E}{\hbar^2} \label{eq:RM-E}
\end{equation}
while $m$ is given by equation~(\ref{eq:parameters-RM1}). The corresponding
$L(m)
= - \alpha^2 m^2 - \alpha^2 \beta^2/(4m^2)$ is a decreasing function of $m$. The
Rosen-Morse problem is therefore of class II. Comparing $\lambda = \lambda_l =
L(l) = - \alpha^2 l^2 - \alpha^2 \beta^2/(4l^2)$ with the expression for
$\lambda$
resulting from equations~(\ref{eq:parameters-RM1}), (\ref{eq:energy-RM}),
and~(\ref{eq:RM-E}), we obtain the relation
\begin{equation}
  l = m - n
\end{equation}
between the eigenvalue labels $n$ and $l$, coming from the resolution of the
Schr\"odinger equation and the factorization method, respectively.\par
%
%-------------------------------------------------------------------
%
Taking equation~(\ref{eq:RM-E}) into account, the operators $S_{\pm}$ of
equation~(\ref{eq:Sbis}) become
\begin{equation}
  S_{\pm} = e^{\pm {\rm i} \xi} \left[\pm {\rm i} \frac{\cosh(\alpha x)}{\alpha}
  \frac{\partial}{\partial x} + \sinh(\alpha x) \frac{\partial}{\partial\xi} +
  \cosh(\alpha x) \frac{\partial}{\partial\eta}\right]
\end{equation}
and the operators $T_{\pm}$ are obtained from them by the transformations $\xi
\leftrightarrow \eta$, $\partial/\partial\xi \leftrightarrow \partial/\partial
\eta$.\par
%
%--------------------------------------------------------------------
%
{}From equations~(\ref{eq:st}), (\ref{eq:s}), and~(\ref{eq:epsilon}) (where
$l+1$ is
replaced by $l$ as we have here a class II problem), we get
\begin{equation}
  s = \frac{|q|}{\alpha l} = \frac{|\beta|}{2l} = - \epsilon a_n \qquad
  t = \epsilon l = \epsilon (m-n) = \epsilon b_n
\end{equation}
with $\epsilon = \beta/|\beta| = B/|B|$. By using
equation~(\ref{eq:psi-RM}) and the
results of reference~\cite{nieto}, the corresponding extended
eigenfunctions can be
written as
\begin{eqnarray}
  \Psi^{(m,\epsilon)}_{s,t}(x,\xi,\eta) & = & (2\pi)^{-1}
N^{(m,\epsilon)}_{s,t} e^{{\rm
         i}(s\xi + t\eta)} e^{-\epsilon s \alpha x} [\cosh(\alpha
x)]^{-\epsilon t}
         \nonumber \\
  && \mbox{} \times {}_2F_1(\epsilon t - m, \epsilon t + m + 1;
\epsilon(t-s) + 1; y)
         \nonumber \\
  && y \equiv \case{1}{2} [1 + \tanh(\alpha x)]
\end{eqnarray}
where
\begin{equation}
  N^{(m,\epsilon)}_{s,t} = \frac{1}{2^{\epsilon t} \Gamma(\epsilon(t-s)+1)}
  \left(\frac{\epsilon \alpha (t-s) (t+s) \Gamma(m+\epsilon t+1)
  \Gamma(m-\epsilon s+1)}{t\, \Gamma(m-\epsilon t+1) \Gamma(m+\epsilon s+1)}
  \right)^{1/2}.
\end{equation}
\par
%
%-------------------------------------------------------------------
%
After some calculations using well-known properties of the hypergeometric
function~\cite{abramowitz}, we obtain
\begin{eqnarray}
  S_{\pm} \Psi^{(m,\epsilon)}_{s,t} & = & \epsilon i \left(\frac{(t-s)
(t+s) (m\mp s)
           (m\pm s+1)}{(t-s\mp1) (t+s\pm1)}\right)^{1/2}
           \Psi^{(m,\epsilon)}_{s\pm1,t} \\
  T_{\pm} \Psi^{(m,\epsilon)}_{s,t} & = & - \epsilon i \left(\frac{(t-s)
(t+s) (t\pm1)
           (m\mp t) (m\pm t+1)}{t\, (t-s\pm1) (t+s\pm1)}\right)^{1/2}
           \Psi^{(m,\epsilon)}_{s,t\pm1}
\end{eqnarray}
which, together with equation~(\ref{eq:S_0-action}), give the action of the
so(2,2)
generators on the extended eigenfunctions of the Rosen-Morse potential. Here the
conserved quantity $m$ is related to the potential parameter $C$. The operators
$S_{\pm}$ change the other potential parameter $B$ into $B' = B (s\pm1)/s$,
while
leaving $n$ fixed, while $T_{\pm}$ change both $B$ and $n$ into $B' = B
(t\pm1)/t$
and $n' = n\mp\epsilon$, respectively.\par
%
%===================================================================
%
\section{The Kepler problem in a space of constant negative curvature}
\setcounter{equation}{0}
In a space of constant negative curvature $-R$, the radial wavefunction for an
electron of mass $\mu$ in a Coulomb potential satisfies the
equation~\cite{infeld45}
\begin{equation}
  \frac{d}{dx} \left(\sinh^2 x \frac{d\psi}{dx}\right) + \left[(\lambda - 2\nu)
  \sinh^2 x + 2\nu \sinh x \cosh x - l(l+1)\right] \psi = 0 \label{eq:H-Kepler}
\end{equation}
where $0 \le x < \infty$, $l$ is the angular momentum,
\begin{equation}
  \nu \equiv \frac{ZR}{a_0} \qquad \lambda \equiv \frac{2\mu R^2}{\hbar^2} E
  \label{eq:nu}
\end{equation}
and $a_0 = \hbar^2/(\mu e^2)$ denotes the Bohr radius. If $R \to \infty$
and $x \to
0$ in such a way that $xR \to r$, then equation~(\ref{eq:H-Kepler}) reduces
to that
of an electron in a central Coulomb field $-Ze^2/r$ in Euclidean space.\par
%
%-----------------------------------------------------------------
%
The negative energy eigenvalues and corresponding eigenfunctions are given
by~\cite{infeld45}
\begin{equation}
  E_n = \frac{Ze^2}{R} - \frac{\hbar^2}{2\mu R^2} \left(n^2 - 1\right) -
  \frac{Z^2 e^4 \mu}{2\hbar^2 n^2} \qquad n = 1, 2, \ldots, n_{max}
\end{equation}
and
\begin{equation}
  \psi_{n_r,l}(x) = N_{n_r,l} \sinh^l x\, e^{\left(n_r - \frac{\nu}{n}\right) x}
  {}_2F_1 \left(-n_r, l+1+\frac{\nu}{n}; 2l+2; w\right) \qquad w \equiv 1 -
e^{-2x}
  \label{eq:psi-Kepler}
\end{equation}
respectively. Here
\begin{equation}
  n = n_r + l + 1 \label{eq:n-Kepler}
\end{equation}
where $n_r$ is the radial quantum number, as in Euclidean space, but now
$n$ only
takes a finite number of values $n_{max}$. The latter corresponds to the
number of
independent functions~(\ref{eq:psi-Kepler}) satisfying the normalization
condition
\begin{equation}
  \int_0^\infty dx\, \sinh^2 x |\psi_{n_r,l}(x)|^2 = 1
\end{equation}
for a given $l$ value, and it is equal to the largest integer smaller than
$\sqrt{\nu}$.\par
%
%------------------------------------------------------------------
%
By setting
\begin{equation}
  \psi(x) = \mbox{\rm cosech} x\, \phi(x)
\end{equation}
equation~(\ref{eq:H-Kepler}) can be rewritten in a form similar to
equation~(\ref{eq:eq-Ebis}) with
\begin{equation}
  p = 0 \qquad \alpha = 1 \qquad q = - \nu
\end{equation}
and $\psi$, $m$, $\lambda$ replaced by $\phi$, $l$, $\lambda - 1 - 2\nu$,
respectively. The function $L(m)$ now becomes $L(l) = - l^2 - \nu^2/l^2$.
It is an
increasing function of $l$. The problem considered is therefore of class I. The
counterpart of $m = 0$, 1, \ldots,~$l$ in the general theory of section~2
is $l=0$,
1, \ldots,~$n-1$, corresponding to $n_r = n-1$, $n-2$, \ldots,~$0$.\par
%
%-------------------------------------------------------------------
%
{}From equations~(\ref{eq:st}), (\ref{eq:s}), and~(\ref{eq:epsilon}) (where
$l$ is
replaced by $n-1$), we get
\begin{equation}
  s = \frac{\nu}{n} \qquad t = -n.
\end{equation}
By using equations~(\ref{eq:psi-Kepler}), (\ref{eq:n-Kepler}), and the
results of
reference~\cite{infeld51}, the corresponding extended eigenfunctions can be
written as
\begin{eqnarray}
  \Psi^{(l)}_{s,t}(x,\xi,\eta) & = & (2\pi)^{-1} N^{(l)}_{s,t}\, e^{{\rm
i}(s\xi + t\eta)}
           \sinh^l x\, e^{-(s+t+l+1)x} {}_2F_1(t+l+1, s+l+1; 2l+2; w)
\nonumber \\
  && w \equiv 1 - e^{-2x}
\end{eqnarray}
where
\begin{equation}
  N^{(l)}_{s,t} = \frac{2^{l+1}}{(2l+1)!} \left(\frac{(s+t) (s-t) (l-t)!\,
\Gamma(s+l+1)}
  {(-t) (-t-l-1)!\, \Gamma(s-l)}\right)^{1/2}.
\end{equation}
\par
%
%-------------------------------------------------------------------
%
When acting on such extended eigenfunctions, the su(1,1) generators $S_0$,
$S_{\pm}$ of equations~(\ref{eq:S_0}) and~(\ref{eq:Sbis}) become
\begin{equation}
  \tilde{S}_0 = - {\rm i} \frac{\partial}{\partial\xi} \qquad
  \tilde{S}_{\pm} = e^{\pm{\rm i}±\xi} \left[\pm \sinh x
\frac{\partial}{\partial x}
  + \cosh x \left( - {\rm i} \frac{\partial}{\partial\xi} \pm 1\right) - {\rm i}
  \sinh x \frac{\partial}{\partial\eta}\right]. \label{eq:Stilde}
\end{equation}
The other su(1,1) generators $T_0$, $T_{\pm}$ of equations~(\ref{eq:S_0})
and~(\ref{eq:Tbis}) are similarly transformed into $\tilde{T}_0$,
$\tilde{T}_{\pm}$, which can be obtained from equation~(\ref{eq:Stilde}) by the
substitutions $\xi \leftrightarrow \eta$, $\partial/\partial\xi
\leftrightarrow \partial/\partial\eta$.\par
%
%--------------------------------------------------------------------
%
After some calculations using well-known properties of the hypergeometric
function~\cite{abramowitz}, we obtain
\begin{eqnarray}
  \tilde{S}_{\pm} \Psi^{(l)}_{s,t} & = & \left(\frac{(s+t) (s-t) (s\mp l)
(s\pm l\pm1)}
         {(s+t\pm1) (s-t\pm1)}\right)^{1/2} \Psi^{(l)}_{s\pm1,t} \\
  \tilde{T}_{\pm} \Psi^{(l)}_{s,t} & = & - \left(\frac{(s+t) (s-t) (-t\mp1)
(-t\pm l)
         (-t\mp l\mp1)}{(-t) (s+t\pm1) (s-t\mp1)}\right)^{1/2}
\Psi^{(l)}_{s,t\pm1}
\end{eqnarray}
which, together with equation~(\ref{eq:S_0-action}), give the action of the
so(2,2)
generators on the extended eigenfunctions of the Kepler problem. Here the
conserved quantity is the angular momentum $l$. The operators $\tilde{S}_{\pm}$
leave $n$ (or $n_r$) unchanged, but change the potential parameter $\nu$
into $\nu'
= \nu (s\pm1)/s$, whereas $\tilde{T}_{\pm}$ change both $n$ (or $n_r$) and $\nu$
into $n' = n \mp 1$ (or $n'_r = n_r \mp 1$) and $\nu' = \nu (t\pm1)/t$,
respectively.
From the definition of $\nu$ in equation~(\ref{eq:nu}), it is clear that
$\tilde{S}_{\pm}$ (resp.\ $\tilde{T}_{\pm}$) relate eigenfunctions of the Kepler
problem in spaces of different curvature, $R$ and $R' = R(s\pm1)/s$ (resp.\
$R' =
R(t\pm1)/t$).\par
%
%==================================================================
%
\section{Conclusion}
In the present paper, we did show that the factorization method can be used
in an
effective way to construct satellite algebras for all the Hamiltonians that
admit a
type E factorization. Special emphasis was laid on the so(2,2) algebras
characterizing hyperbolic potentials, but it is clear that a similar
analysis could be
carried out for the so(2,1) algebras appropriate to trigonometric potentials.\par
%
%--------------------------------------------------------------------
%
In the examples considered, we found that the conserved quantity, which is the
eigenvalue of the satellite algebra Casimir operator, may have various physical
meanings: a combination of the potential parameters in the GMP case, one of the
two potential parameters for the Rosen-Morse potential, and the angular momentum
quantum number in the Kepler problem in a space of constant negative curvature.
Similarly, the algebra generators may have various physical effects: relating
eigenfunctions belonging to different potentials of the same family in the
first two
cases, or connecting eigenfunctions in spaces of different curvature in the
last one.
This hints at the existence of physical applications that may have been
overlooked
so far.\par
%
%-------------------------------------------------------------------
%
The approach used in the present paper is not the only one allowing the
construction
of satellite algebras or, more generally, providing an algebraic treatment
of the
problems considered. Of particular significance is the work of Wu {\em et
al}~\cite{wu}, who determined an so(2,2) algebra for the class of Natanzon
potentials~\cite{natanzon}, which includes all the potentials solvable in
terms of
the hypergeometric or confluent hypergeometric function. A treatment of the same
in terms of an so(2,1) algebra was also given by Cordero and
Salam\'o~\cite{cordero}. It is worth mentioning too that the Kepler problem in a
space of constant negative curvature was analysed in terms of a quadratic
algebra~\cite{granovskii}. We would like to stress however that our
approach is the
only one establishing a clear link with the factorization method of
Schr\"odinger~\cite{schrodinger}, and Infeld and Hull~\cite{infeld51}, and
that in
comparison with other papers we show more explicitly in the examples the effect
of the action of the satellite algebra generators in terms of the potential
parameters and the energy, what could be important from a physical viewpoint. A
more detailed mathematical discussion of the irreducible representations will be
given elsewhere.\par
%
%--------------------------------------------------------------------
%
As mentioned in section~1, considering type A and E factorizations is not a
restriction as the other factorization types are but limiting cases of them. In
forthcoming publications, we hope to come back to such limiting cases, as
well as
to a generalization of the factorization method recently proposed by Cari\~ nena
and Ramos~\cite{carinena}.\par
%
%====================================================================
%

\end{document}